\documentclass[12pt]{article}
\usepackage[russian,english]{babel}
\usepackage[cp1251]{inputenc} 
\usepackage{graphicx}
\usepackage[nooneline]{caption} 
\begin{document}

%%% Title section
\begin{center}
{\Large\bf Applying statistical methods to text steganography}
\end{center}
\begin{center}
{\sc Nechta I., Fionov A.}\\
{\it Siberian State University of Telecommunication and Information Sciences \\
Novosibirsk, Russia\\}
e-mail: {\tt www@inbox.ru}
\end{center}

%%% abstract
\begin{abstract}
This paper presents a survey of text steganography methods used for hiding secret information inside some covertext. Widely known hiding techniques (such as translation based steganography, text generating and syntactic embedding) and detection are considered. It is shown that statistical analysis has an important role in text steganalysis.

\end{abstract}

%%%
\section*{Introduction}

Steganography is the art and science of writing hidden messages in such a way that no one, apart from the sender and intended recipient, suspects the existence of the message. In steganography, it is very important to find a good covertext suitable for embedding hidden messages. This paper provides a basic introduction to steganography and steganalysis, with a particular focus on text steganography. Information hiding techniques are discussed, providing motivation for moving toward text steganography and steganalysis.We will show some of the problems inherent in text steganography as well as issues with existing solutions.

%%%
\section{Steganography}

In 1984, Gustavus Simmons illustrated what is now widely known as the prisoners' problem: Let us consider Fig1. two accomplices in a crime, Alice and Bob, are arrested in separate cells. They want to coordinate an escape plan, but their only means of communication is by way of messages conveyed for them by Wendy the warden. Should Alice and Bob try to exchange messages that are not completely open to Wendy, or ones that seem suspicious to her, they will be put into a high security prison. Alice and Bob will have to deceive the warden by finding a way of communicating secretly in the exchanges. It can be done such way: Alice gets any text (covertext) which does not arise the warden suspicion and embeds (using steganographic method) secret message into it. Then she sends the covertext with message to Bob. This covertext is available to both Warden and Bob, but it contains different information to Wendy than to Bob.

\begin{figure}[htbp]
\begin{center}
\includegraphics[width=12cm]{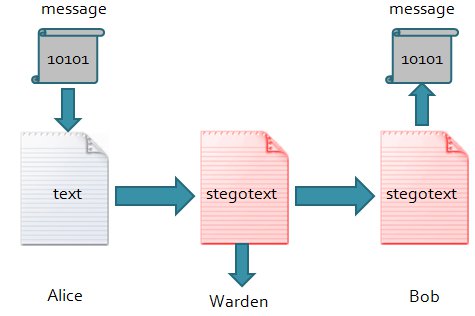}
\caption{Basic steganography protocol}
\end{center}
\end{figure}

Many types of covertexts are based on data having redundancy, such as video, audio, or image files. In this article we'll discuss one of the areas of steganography, which uses text files as a covertext. A file with covertext is called container. 

%%%%%%%%%%%%%%%%%%%%%%%%%%%%%%%%%%%%%%%%%%%%%%%%%%%%
\section{Information hiding methods}

Existing methods of embedding secret messages in the text data could be divided into three groups:
\begin{enumerate}
\item \textbf{ Syntactic methods}
The early methods of text information hiding are based on the physical formatting of text. One of such methods, for example, proposed in \cite{PPF}, uses the extra space between words. One space means that the transmitted information bit is ``0'', and two spaces mean ``1''. This technique is widely used in HTML files (web pages) because space presence does not affect on the web page appearance.The disadvantage of this method is easy detectability, extra spaces are not used in text. It is possible to use special characters instead of spaces wich do not appear in commonly used text editors.

Another method proposed in \cite{PPF} uses a syntax error when writing words such as:
\newline
\textit {``This is the end''}
\newline
\textit {``This iz the end''}
\newline 
The second version has a misprint. The presence of errors in certain words (in particular "is") means that the transmitted information bit is ``0'', and errors absence  means that bit is ``1'' . Thus, there is a transfer of information in the text. This method is not easily detectable, because some errors may occur in the message.
\item \textbf{Semantic methods}
This group includes Tyrannosaurus Lex(T-Lex), published in \cite{Tlex}, which uses the replacement words in the sentence on their synonyms, for example:

\begin{table}[tbph]
\begin{center}
\begin{tabular}{c|c|c|c|}
\multicolumn{1}{c}{}& \multicolumn{1}{c}{ \textbf{excellent}}  &  \multicolumn{1}{ c}{}&  \multicolumn{1}{c}{\textbf{city}}\\
 & (0) decent & & (0)metropolis \\
Tobolsk is a & &little & \\
 &(1) fine & & (1) town\\
\end{tabular}
\end{center}
\end{table}

Message embeded by synonym selection. Sentence \textit{``Tobolsk is a decent little town''} contains message -  \textit{``01''}. This method requires a large synonyms dictionary. The below examples illustrate two shortcomings of the T-Lex system. First, it sometimes replaces words with synonyms that do not agree with correct English usage, as seen in the phrase \textit{``soon subsequently dispatched''}. Second, T-Lex also substitutes synonyms that do not agree with the genre and the author style of the given text.

 \textit{An invitation to dinner was soon afterwards dispatched}
\newline
 \textit{An invitation to dinner was soon subsequently dispatched}

 \textit{\ldots and make it still better, and say nothing of the bad belongs to you alone.}
\newline
 \textit{\ldots and make it still better, and say nada of the bad belongs to you alone.}

It is clear that the word  \textit{``nada''} does not belong to Jane Austen's style. Furthermore, the string  \textit{``say nada''} of is not part of typical English usage.

There is another approach, proposed in ~\cite{paraphrase}, of generating sentence level paraphrases for information hiding. Example: 
\newline
\textit{The caller identified the bomber as Yussef Attala, 20, from the Balata refugee camp near Nablus.
 \newline
The caller named the bomber as 20-year old Yussef Attala from the Balata refugee camp near Nablus.}
\newline
This method has a high degree of secrecy.

\item \textbf{ Linguistically-driven generation methods}

Let's consider the method proposed in ~\cite{chapman}, using a context-free grammar to generate a natural like text.

\textbf{Grammar Rules:}
\newline
${S} \rightarrow{ABC}$
\newline
${A} \rightarrow {She(0) \mid He(1)}$
\newline
${B} \rightarrow {likes(0) \mid hates(1)}$
\newline
${C} \rightarrow {milk(0) \mid apples(1)}$

This approach produces stegotext that looks similar to the real structure of the original text. It is used a set of grammatical rules to generate stegotext and the choice of each word determines how secret message bits are encoded. The quality of the resulting stegotext directly depends on the quality of the grammar. Today's most popular stegosystems are Nicetext ~\cite{nicetext}, Texto ~\cite{texto} and Markov-Chain-Based ~\cite{mcb}, because they have high ratio of the input message size to the generated text size. Also, resulting stegotext looking like natural text but it should be noted that, as usually, such text is meaningless.

The next one approach was proposed at ~\cite{tbs}. The key idea is to hide information in a noise than occurs invariably in natural languages transformation. When translation a non-trivial text between a pair of natural languages, there are typically many possible translations. Selecting one of these translations can be used to encode information. For example:

\textit{``\textcyrillic {Джек нанес краску на стену}''} can be translated as: 
\newline
\textit{``Jack spayed paint on the wall''}, or \textit{``Jack sprayed the wall with paint''}.
\end{enumerate}

\section{Steganalysis methods}

There converse problem of steganography is steganalysis. Its goal is to identify suspected container, determine whether or not they have embedded message in it, and, if possible, recover that message. Statistical attacks are commonly used for stegotext detection. For example, widely known support vector machines (SVMs) ~\cite{libsvm} are a set of related supervised learning methods that analyze data and recognize patterns, used for classification. There are two types of errors uses for evaluating the steganalysis methods reliability:
\newline
\textit{False Positive} errors occur when the method mistakenly flags an natural text as stegotext.
\newline
\textit{False Negative} errors occur when the method mistakenly flags stegotext as natural text.

The most easily detectable methods are syntactic because they could be detected by simple analyzer. Presence of double spaces in text might cause suspicion.
It was noted earlier that methods of natural-like text generation have one disadvantage --- resulting text is meaningless. It  requires a human intervention to determine the meaningfulness of the text. However, it is not always possible, because of the large volume of messages transmitted in the network. It is necessary to create automated methods for steganalysis. Nowadays, there are a large number of different steganalysis methods.Let us consider in more detail the following method. A method using semantic shortcomings of methods published in ~\cite{topkara}. When you replace the words on their synonyms can break semantic rules, for example: 

\textit{``What time is it ?''} Word \textit{``time''} could be replaced as \textit{``period''} or \textit{``duration''}that do not agree with correct English usage. False positive ratio is 38.6\%. False negative - 15.1\%. Low reliability level makes it difficult to the practical application of this method. In addition, it is requires a lot of time working, and a large database of language rules. 
Method, proposed in \cite{zchen}, uses word frequency and its variance in the analyzed text. Obtained data and The Support Vector Machine (SVM) used for identify stegotext Nicetext, Texto or Markov-Chain-Based presence when container size more than 5Kb. Sum of errors less than 7.05\%.

The most effective steganalysis for Nicetext stegosystem proposed in ~\cite{n400}. In virtue of the concepts in area of information theory, the method uses an information entropy-like statistical variable of words in detected text segment together with its variance as two classification features for SVM. The method was centered on detection for small size text segments estimated in the hundreds in words. The experimental accuracy of the method on classification of generated text and normal text exceeds 99\% when text size is larger than 400 bytes. Even for sentences, the experimental accuracy exceeds 85\%.

New method of statistical analysis was suggested in ~\cite{my}. The compression used for stegotext detection. It is known that an embedding message breaks statistical structure of the container, increasing its entropy. Consequently, the full container will compress worse than empty. Let us consider the example:

A and B are empty and full containers, respectively.
\begin{table}[tbph]
\caption{Container size before and after compression}
\begin{center}
\begin{tabular}{|c|c|c|c|}
\hline
\multicolumn{1}{|c|}{container}& \multicolumn{1}{|c|}{ before compr.}  &  \multicolumn{1}{| c|}{after compr.}\\
 \hline
A&500&320\\
\hline
B&500&300\\
\hline
\end{tabular}
\end{center}
\end{table}

Add content of suspected container C into A and B. Compare added content sizes before and after compression. 

\begin{table}[tbph]
\caption{Content size of C}
\begin{center}
\begin{tabular}{|c|c|c|c|}
\hline
\multicolumn{1}{|c|}{container}& \multicolumn{1}{|c|}{ before compr.}  &  \multicolumn{1}{| c|}{after compr.}\\
 \hline
C&50&45\\
\hline
C&50&20\\
\hline
\end{tabular}
\end{center}
\end{table}

It could be asserted that container C is statistically depend with B, which ensures good compression. This principle used in attack on Texto. Accuracy of detection exceeds 99.98\% when text size is larger than 400 bytes.

The next one ~\cite{china}, statistical method used for attack on stegotext Nicetext, Texto, and Markov-Chain-Based. The average length of words, frequency of spaces, letter distribution of the words, first letter of word distribution used as features in SVM classificator. Detection accuracy exceeds 84.42\% for text segments larger than 500 bytes.

\begin{table}[tbph]
\caption{The effectiveness of existing steganalysis methods}
\begin{center}
\begin{tabular}{|c|c|c|c|}
\hline
\multicolumn{1}{|c|}{stegosystem}& \multicolumn{1}{|c|}{ 400 bytes}  &  \multicolumn{1}{| c|}{1 Kb}&  \multicolumn{1}{| c|}{5 Kb}\\
 \hline
Nicetext&99.61\%&99.61\%&99.61\%\\
\hline
Texto&99.98\%&99.98\%&99.98\%\\
\hline
MCB&84.42\%&87.61\%&99.46\%\\
\hline
\end{tabular}
\end{center}
\end{table}

Analysis of translation-based steganography was published in ~\cite{meng}. This method has improved on a previously proposed linguistic steganalysis method based on word distribution which is targeted for the detection of linguistic steganography like nicetext and texto. The newmethod aims to detect the application of TBS and uses none of the related information about TBS, its only used resource is a word frequency dictionary obtained from a large corpus, or a so called natural frequency dictionary, so it is totally blind. It is known, that stegotext consist less high frequency word then in natural text or translated text. This Method needs to know the machine translator set and covertext language. The experimental results show that the method accuracy is 87.7\% when container size - 20 Kb.

The most effective method was proposed in ~\cite{blind}. It is used two features (words frequency and variance of the words distances as sentences structures) for SVM classification.

 Let us consider distances of words in \textit{``What is a web browser?''}.

\begin{table}[tbph]
\caption{Word distances}
\begin{center}
\begin{tabular}{|c|c|c|c|}
\hline
\multicolumn{1}{|c|}{}& \multicolumn{1}{|c|}{ what}  &  \multicolumn{1}{| c|}{web}&  \multicolumn{1}{| c|}{browser}\\
 \hline
what&0&1&2 \\
\hline
web&1&0&1\\
\hline
browser&2&1&0\\
\hline
\end{tabular}
\end{center}
\end{table}

It should be noted than the sentences structures of stegotext looks more ``noisy'' than natural text. The total detection accuracy are 97.65\%, 98.88\% and 99.69\% respectively when the text size is 10Kb, 15Kb and 20 Kb.
\section*{Conclusion}
This paper presents a background on the major algorithms of text steganography and steganalysis. It is shown that overwhelming majority of effective steganalysis methods are based on statistical analysis. Most existing embedding methods could be detected with high probability.

\end{document}